# STRUCTURAL ANALYSIS OF SUPERCONDUCTING ACCELERATOR CAVITIES *

D. Schrage, LANL, Los Alamos, N.M., USA

*Abstract*

The static and dynamic structural behavior of superconducting cavities for various projects was determined by finite element structural analysis. The $\beta$ = 0.61 cavity shape for the Neutron Science Project was studied in detail and found to meet all design requirements if fabricated from five millimeter thick material with a single annular stiffener. This 600 MHz cavity will have a Lorentz coefficient of $-1.8$ Hz/(Mv/meter)$^2$ and a lowest structural resonance of more than 100 Hz.

Cavities at $\beta$ = 0.48, 0.61, and 0.77 were analyzed for a Neutron Science Project concept which would incorporate 7-cell cavities. The medium and high beta cavities were found to meet all criteria but it was not possible to generate a $\beta$ = 0.48 cavity with a Lorentz coefficient of less than $-3$ Hz/(Mv/meter)$^2$.

## 1 INTRODUCTION

There are quite a few accelerator projects underway for which elliptical superconducting cavities are planned. This paper documents structural analysis of $\beta < 1$ superconducting cavities for the Neutron Science Project of JAERI [1] and includes consideration of Lorentz force detuning, cavity fabrication, vacuum loading, tuning forces, and mechanical resonant frequencies.

## 2 TECHNICAL CONSIDERATIONS

While each accelerator has specific technical requirements with regard to the values of $\beta$, the number of cells, and the bore sizes of the cavities, there are some other physics and engineering considerations that must be included in the design of the cavities. Some of these parameters, the peak electric and magnetic surface fields along with the bore radius, affect the performance of the cavities. Other parameters, such as the material thickness and the wall slope, are related to the practical matter of manufacture of the cavities. Lastly, the presence or absence of annular stiffeners has a significant effect upon the Lorentz for RF detuning, the mechanical resonant frequencies, and the tuning forces. A detailed discussion of this is given in References 2 and 3.

The parameters are listed in Table 1. The value selected for $B_{peak}/E_{peak}$ is arguable and some organisations would suggest that a higher value would be more suitable if it resulted in a lower peak electric field. Indeed, all of the values are to some extent arbitrary; they are certainly not absolute. However, they do serve as guidelines for preliminary design of cavities.

Table 1: Cavity Design Parameters

| PARAMETER | ALLOWABLE VALUES |
|---|---|
| **Peak Electric Field** | $E_{peak}/E_a$ = minimum |
| **Peak Magnetic Field** | $B_{peak}/E_{peak}$ ~ 1.71 mT/(Mv/meter) |
| **Fabrication** | $R_{min}$ > 2*thickness |
| **BCP Cleaning** | Slope $\geq 6°$ |
| **Mech. Resonances** | $\omega_1$ > 60 Hz |
| **Radiation Pressure** | $|k|$ < 2.0 Hz/(Mv/meter)$^2$ |
| **Tuning Sensitivity** | < 5.0 #/kHz |
| **Vacuum Loading** | $\sigma_{von-Mises}$ < 3,500 #/in$^2$ |

## 3 STATIC ANALYSIS OF CAVITIES

Three cavity mid-cell shapes were analysed: $\beta$ = 0.48, 0.61, and 0.77. These were obtained from Reference 4 and are shown on Figure 1. A $\beta$ = 1 cross-section is shown for reference.

The structural analysis was carried out using COSMOS/M™[5]. Two-dimensional axi-symmetric elements were used for the analysis of half-cells to determine the tuning forces plus the deflections, stresses and frequency shifts under vacuum load and Lorentz pressure. The frequency shifts were determined from the output of SUPERFISH [6].

The main consideration was the Lorentz force detuning. The analyses were performed for various stiffener ring radii. The results for the $\beta$ = 0.61 cavity are shown on Figure 2 for material thicknesses of 3, 4, and 5 mm. The results are similar for the other two cavities. Without annular stiffeners none of the cavity shapes will satisfy the requirement that the Lorentz detuning coefficient of the cavity be less than $-3$ Hz/(Mv/meter)$^2$. However, for the $\beta$ = 0.61 with the 4 mm thickness, the curve is quite flat so the selection of the 7 inch stiffener radius is not rigid.

Some cases were run with two stiffener rings but these resulted in unacceptably high tuning forces. Use of two stiffener rings would also increase fabrication costs.

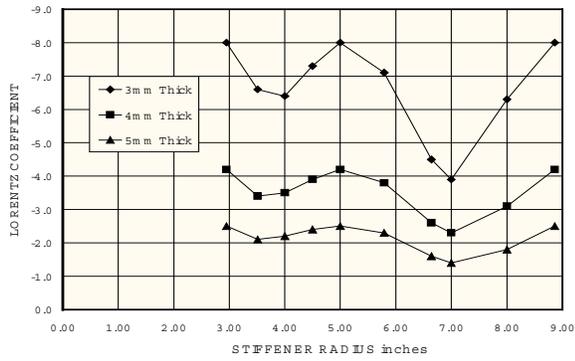

Figure 2: Effect of Stiffener Radius and Material Thickness for β = 0.61 Cavity

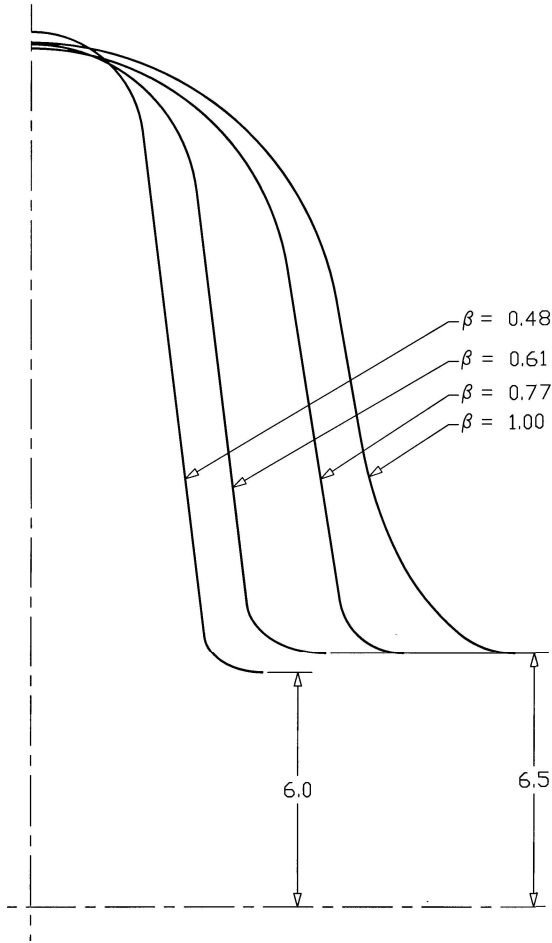

Figure 1: Cavity Cross-Sections

The results for the three cavities are listed on Table 2. The β = 0.48 cavity does not meet the fabrication criteria ($R_{min} > 2.0*t$) and has a Lorentz coefficient that is greater than the specified value. However, with the lower-β cavities operated at lower gradient (the requirement is that $E_{peak} < 16.0$ Mvolt/meter [1]) this may be acceptable. At this peak electric field, the accelerating field is only 3.7 Mvolt/meter and the Lorentz detuning is reduced to 1/8th the value at $E_a = 10$ Mvolt/meter.

Table 2: Static Analysis Results for Stiffened Cavities

|  | β = 0.48 | β = 0.61 | β = 0.77 |
|---|---|---|---|
| **Thickness, mm** | 5.0 | 4.0 | 4.0 |
| **$R_{min}$** | 1.4*t | 4.0*t | 3.5*t |
| **k Mvolt/m²** | -3.3 | -1.8 | -0.9 |
| **Tuning #/kHz** | 1.06 | 0.97 | 1.78 |
| **Vac. Stress #/in²** | 3496 | 3811 | 2896 |

The deformation of the β = 0.48 cavity under Lorentz pressure resulting from an accelerating gradient of 10 Mvolt/meter is shown on Figure 3. The Lorentz pressures are quite low with the maximum being 0.48 #/in². The axial deformations are similarly low; the maximum is 6.7 X $10^{-6}$ inch. This corresponds to a frequency shift of –330 Hz.

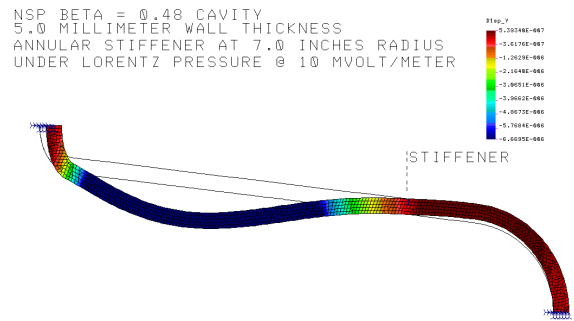

Figure 3: Lorentz Pressure Deformation for β = 0.48 Cavity

Three-dimensional finite element models were used to determine the gravity deformations of the complete 5-cell and 7-cell cavities. These analyses were run using COSMOS/M with three-node shell elements. The weights and mid-length transverse deflections of the cavities are listed on Tables 3 and 4. The presence of the stiffeners produces a significant reduction of the deflection.

Table 3: Static Deflections of 7-Cell Cavities

|  | β = 0.48<br>5 mm Thick | β = 0.61<br>4 mm Thick | β = 0.77<br>4 mm Thick |
|---|---|---|---|
| **Un-Stiffened** | | | |
| **Wt (#)** | 237. | 195. | 211. |
| **Disp. (in)** | 0.01474 | 0.01537 | 0.02854 |
| **Stiffened** | | | |
| **Wt (#)** | 278. | 232. | 251. |
| **Disp. (in)** | 0.00059 | 0.00047 | 0.00061 |

Table 4: Static Deflections of 5-Cell Cavities

|  | β = 0.48<br>5 mm Thick | β = 0.61<br>4 mm Thick | β = 0.77<br>4 mm Thick |
|---|---|---|---|
| **Un-Stiffened** | | | |
| **Wt (#)** | 169. | 139. | 151. |
| **Disp. (in)** | 0.00395 | 0.00622 | 0.00778 |
| **Stiffened** | | | |
| **Wt (#)** | 199. | 166. | 179. |
| **Disp. (in)** | 0.00017 | 0.00020 | 0.00019 |

## 4 DYNAMIC ANALYSIS OF CAVITIES

The three-dimensional finite element models described in the previous paragraph were used to determine the mechanical resonant frequencies. A cross-section of a 5-cell, un-stiffened β = 0.61 cavity is shown on Figure 5 and the results for 5-cell and 7-cell cavities are listed on Tables 5 and 6 respectively. For these cases, the irises of the end-cells were held rigidly fixed in all coordinates. Use of other boundary conditions would have resulted in lower frequencies.

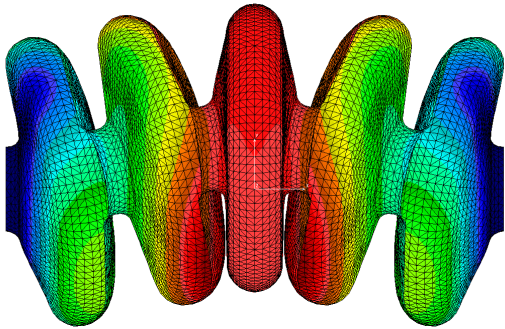

Figure 5: Lowest Mode of β = 0.61 5-Cell Cavity

Table 5: Cavity Structural Frequencies of 5-Cell Cavities

| CAVITY | WALL THICK mm | UN-STIFFENED CAVITY LOWEST FREQUENCY Hz | STIFFENED CAVITY LOWEST FREQUENCY Hz |
|---|---|---|---|
| β = 0.48 | 5.0 | 47. | 181. |
| β = 0.61 | 4.0 | 40. | 217. |
| β = 0.77 | 4.0 | 37. | 251. |

Table 6: Cavity Structural Frequencies of 7-Cell Cavities

| CAVITY | WALL THICK mm | UN-STIFFENED CAVITY LOWEST FREQUENCY Hz | STIFFENED CAVITY LOWEST FREQUENCY Hz |
|---|---|---|---|
| β = 0.48 | 5.0 | 27. | 130. |
| β = 0.61 | 4.0 | 22. | 130. |
| β = 0.77 | 4.0 | 20. | 142. |

Past experiments [7] have shown good agreement of measured mechanical resonant frequencies with the predicted values. It is important to note that the analyses were run for simple cavities; there were no beam tubes, power couplers, HOM couplers, etc. included. In addition, there is no consideration of the stiffness of the cavity support structure. Inclusion of any or all of these items will reduce the mechanical resonant frequencies. Thus, the frequencies listed in Tables 5 and 6 must be regarded as ideal maximums. As in the case of a similar study of the cavities for the APT linac [8], it was found that the annular stiffeners would be required to meet the dynamic requirements, in particular when the effects of the beam tubes, etc. are included.

## 5 CONCLUSIONS

There are many variables to consider in the design of superconducting cavities. However, in meeting the requirements listed in Table 1, the options diminish rapidly. It is clear that for values of β < 0.5, the structural design of these cavities is a challenge at 600 MHz. Minimization of the Lorentz force detuning will likely require operation of β < 0.5 cavities at $E_a$ < 10 Mvolt/meter. It is also clear that stiffeners will be required to meet the mechanical resonant frequency requirement.

## 6 ACKNOWLEDGEMENT

Rick Wood provided the software support for the calculation of the frequency shifts. Jim Billen and Frank Krawczyk provided cavity designs and SUPERFISH runs for these analyses.